\documentclass[12pt,twocolumn]{article}
\usepackage[a4paper,left=20mm,right=20mm,top=25mm,bottom=25mm,includeheadfoot]{geometry}

\usepackage{mathpazo}
\usepackage{graphicx}
\usepackage{amsmath,textcomp}
\usepackage{makeidx}
\makeindex

\usepackage{sectsty}
	\allsectionsfont{\sffamily\raggedright}
	\sectionfont{\sffamily\large\raggedright}
	\subsectionfont{\sffamily\normalsize\raggedright}
\usepackage{ftnright}

\usepackage{widetext}
\usepackage{flushend}
\usepackage{cuted}
\usepackage{color}
\usepackage{graphicx}
\usepackage{hyperref}
\usepackage{pstricks,amssymb}

\usepackage{fancyhdr}
\begin{document}
\title{\vspace{-2em}\bfseries\sffamily Motion of a rod pushed at one point in a weightless environment in space}
\author{\normalsize Ashok K. Singal\\[2ex]
Astronomy and Astrophysics Division, Physical Research Laboratory\\
Navrangpura, Ahmedabad 380 009, India.\\
{\tt ashokkumar.singal@gmail.com}
}
\date{\itshape Submitted on xx-xxx-xxxx}
\maketitle

\thispagestyle{fancy}


\begin{abstract}
{\sffamily
We analyze the motion of a rod floating in a weightless environment in space when a force is applied at some point on the rod in a direction perpendicular to its length. If the force applied is at the centre of mass, then the rod gets a linear motion perpendicular to its length. However, if the same force is applied at a point other than the centre of mass, say, near one end of the rod, thereby giving rise to a torque, then there will also be a rotation of the rod about its centre of mass, in addition to the motion of the centre of mass itself. If the force applied is for a very short duration, but imparting nevertheless a finite impulse, like in a sudden (quick) hit at one end of the rod, then the centre of mass will move with a constant linear speed and superimposed on it will be a rotation of the rod with constant angular speed about the centre of mass. However, if force is applied continuously, say by strapping a tiny rocket at one end of the rod, then the rod will spin faster and faster about the centre of mass, with angular speed increasing linearly with time. As the direction of the applied force, as seen by an external (inertial) observer, will be changing continuously with the rotation of the rod, the acceleration of the centre of mass would also be not in one fixed direction. However, it turns out that the locus of the velocity vector of the centre of mass will describe a Cornu spiral, with the velocity vector reaching a final constant value with time. The mean motion of the centre of mass will be in a straight line, with superposed initial oscillations that soon die down.
}\\ 
\hrule
\end{abstract}
\section{Introduction}
Consider a uniform rod of length $l$ and mass $m$, freely floating in space in a weightless condition. Suppose a force $f$ is applied at some point on the rod, in a direction perpendicular to the length of the rod. What will be the motion of the rod? The question whether any such rod in space, when pushed at say, one end of the rod, will have only a linear motion or have only a rotation or possess both, has been argued in various forums on the web \cite{n1,n2,n3,n4}. This is a problem involving linear momentum of the centre of mass as well as moment of inertia, angular momentum and rotation about the centre of mass of the system. If the force is applied at the centre of mass $C$ of the rod, then from the law of conservation of momentum, the rod would gain a linear motion in the direction of the applied force. However, if the same force is applied at a point other than the centre of mass, then in addition to the motion of the centre of mass as before, there could also be a rotation of the rod about $C$, due to a finite torque \cite{ki65,da13,go68}. If the force is applied continuously, then any rotation of the rod would imply a continuous change in the direction of the applied force and a consequential change in the direction of acceleration of the centre of mass. One expects the combined motion to be quite complicated. 

We could imagine the rod (as well as the observer) to be freely floating in space, say, in a weightless environment within a satellite orbiting the Earth (and thus freely falling in Earth's gravitational field) which can then be considered to be an inertial frame, provided any tidal effects over the system dimensions due to Earth's gravitation field could be ignored. We take the length of the rod to be short enough so that we may not be bothered about any light-travel time effects. All movements are also assumed to be slow enough so that no special relativistic effects come into picture. Even the sound speed within the rod, with which one part of the rod material may communicate with other parts, i.e., the speed with which any influence within the rod may travel, is taken to be 
fast compared to any translational or rotational speeds of the rod for whatever temporal intervals we may be concerned with.
In order to provide a continuous force perpendicular to the length of the rod, we could strap a tiny rocket to the rod at a point of our choosing. 
We further suppose that the rocket system, providing the thrust, makes only an imperceptible, if any, change in the mass of the rod. Further, we take the force, acceleration etc., though perpendicular to the rod, but to be always in the $x$-$y$ plane, so that the torque, angular momentum and angular velocity vectors will all be along the $z$-axis, therefore we need to consider only the magnitudes of such vectors and as we shall see, it does not give rise to any ambiguities.

\section{An impulse given to one of a pair of independent masses}
\begin{figure*}
\includegraphics[width=\linewidth]{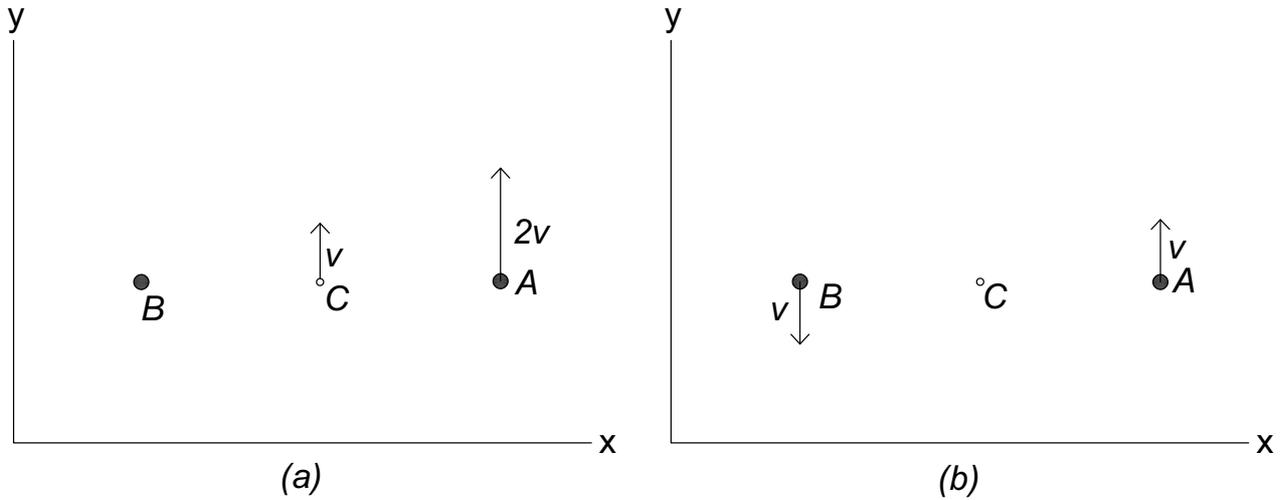}
\caption{({\bf a}) Particle $A$, with mass $m/2$, is given an impulse $p$, so that it moves with a constant velocity $v_A=2p/m$ along the $y$ direction. The centre of mass $C$ then moves with a linear velocity $v=p/m$ in $y$ direction. ({\bf b}) With respect to $C$, $A$ is moving with a speed ${v}$ along the $y$ direction while $B$ is moving with the same speed ${v}$ in the opposite direction, implying that the system comprising particles $A$ and $B$ possesses an angular momentum $mlv/2$ about its centre of mass $C$.}
\end{figure*}
In a composite system, comprising two or more particles whose relative coordinates may or may not be governed by any constraints between them, any net force which is not passing through the centre of mass of the composite system, will create torque and thereby impart angular momentum to the system about its centre of mass \cite{da13}.

In order to demonstrate that a linear motion of the centre of mass alone might not suffice to describe the dynamics of a system even in a weightless environment in space (vacuum!) we first consider a case where the mass of the system is in the form of two independent, equal point masses. Let two particles, $A$ and $B$, each of mass $m/2$, lie initially a distance $l$ apart, parallel to the $x$-axis. Let us now give a push to $A$, say along the y-axis, i.e., in a direction perpendicular to its separation from $B$. For this we consider a force ${f}$ along $y$-axis, applied for a short (infinitesimal!) duration $\Delta t$, nevertheless imparting a finite impulse $p={f} \Delta t$, like in a sudden (quick) hit on particle $A$. As a consequence of the impulse given, particle $A$ with mass $m/2$ gains a velocity ${v}_A=2 {p}/m$ along $y$ direction, having a kinetic energy $K=p^2/m$, which is the total kinetic energy of the system, since $B$ is stationary. From the conservation of momentum, the centre of mass $C$ of the system moves with a velocity ${v}={p}/m$ parallel to the $y$-axis (Fig. 1a), with a kinetic energy of translation 
\begin{equation}
\label{eq:73a1}
K_1=p^2/2m.
\end{equation}
But this accounts for only half of the total kinetic energy of the system, i.e., $K_1=K/2$. 

Actually, the system possesses additional motion apart from the linear motion of its centre of mass. This is because while $A$ has a velocity $2v$ along $y$ direction, $B$ is stationary,  and $C$ has a velocity $v$ along $y$ direction. Therefore 
with respect to $C$, $A$ has a speed ${v}$ along $y$ direction while $B$ has a speed ${v}$ in the opposite direction (Fig.~1b) and these two together constitute an angular momentum 
\begin{equation}
\label{eq:73a2}
J=2(m/2) (l/2)v=mlv/2 
\end{equation}
about $C$.

The kinetic energy associated with the motion about $C$ is 
\begin{equation}
\label{eq:73a3}
K_2=2(m/2)v^2/2=p^2/2m, 
\end{equation}
which yields $K_1+K_2=K$. 

Thus, when a system is given a push, the liner motion of the centre of mass alone may not describe the total dynamics of the system, which might as well, in addition, possess a motion about its centre of mass, irrespective of whether the system is on Earth or in space.
\section{A push given to a rod with a uniform distribution of mass}
\subsection{Force applied for a short duration}
A rod with a uniform distribution of mass $m$ and length $l$ has a moment of inertia about its centre of mass $C$ as $I = m l^2/12$ \cite{ki65,da13,go68}. Let us suppose that the rod, to begin with, is lying along the $x$-axis and we apply, at a distance $\delta$ from $C$, a force $f$ along the $y$-axis, i.e., in a direction perpendicular to the length of the rod, for a short duration $\Delta t$, imparting a finite impulse, $p={f} \Delta t$ to the system, like in a sudden (quick) hit at one end of the rod. As a result, the centre of mass $C$ of the system moves with a velocity ${v}={p}/m$ along $y$ direction, with a kinetic energy of translation $K_1=m v^2/2=p^2/2m$. But in addition there is a torque $N= f \delta$, about $C$, for time $\Delta t$ that gives rise to an angular momentum  $J=I \omega =m l^2\omega/12=N \Delta t= p \delta$, about $C$. From this we can readily see that the rod would rotate around $C$ with an angular speed $\omega =12 p \delta/ m l^2$. The kinetic energy of rotation about $C$ would be $K_2=I \omega^2/2=6 p^2 \delta^2/m l^2$, with total kinetic energy of the system being 
\begin{equation}
\label{eq:73a3.1}
K=K_1+K_2=(p^2/2m)(1 + 12 \delta^2 /l^2).
\end{equation}

If the force is applied at one end of the rod, then $\delta=l/2$ and $\omega =6 p/ m l$. The kinetic energy of rotation will then be three times that of translation, with total kinetic energy $k=2p^2/m$. The centre of mass $C$ of the rod will be moving along the direction of the given impulse with a constant velocity $v=p/m$ with the rod simultaneously 
spinning about $C$ with an angular frequency $\omega =6 v/ l$ (Fig. 2). 
\begin{figure}[t]
\begin{center}
\includegraphics[width=\columnwidth]{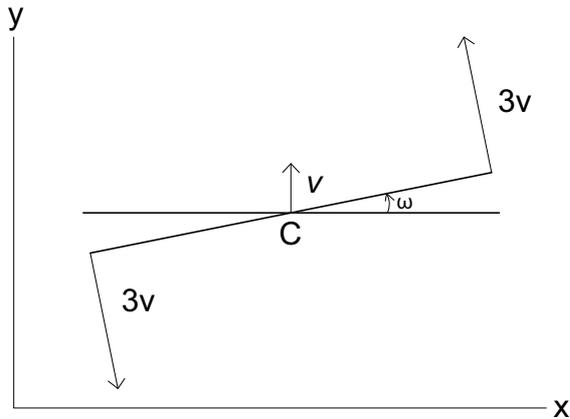}
\end{center}
\caption{After a force is applied for a short duration at one end of the rod, the centre of mass $C$ of the rod moves with a constant linear velocity $v$ along the $y$ direction. In addition,  the rod rotates with a constant angular speed $\omega = 6v/l$ about $C$.} 
\end{figure}

Of course, if the force is applied at the centre of mass $C$, then $\delta=0$ and $\omega =0$, i.e., the rod does not rotate and from Eq.~(\ref{eq:73a3.1}) the total kinetic energy of the system is $K=K1=(p^2/2m)$.
\subsection{The force applied continuously}
If a finite force $f$ is applied continuously at a distance $\delta$ from $C$, in a direction perpendicular to the rod, then $C$ gets accelerated at a rate $a=f/m$, while due to the torque $N= f \delta$, there is an increasing angular momentum, $\dot {J}=N$ or $I \dot\omega = f \delta$, implying an angular acceleration, $\dot\omega=f \delta/I$, about $C$. 
The rod will rotate about $C$ with an angular speed of rotation $\omega =t f \delta/ I$ and  a rotation angle, $\phi= t^2 f \delta/2I$, assuming $\phi=0$ at $t=0$. 

With the rotation of the rod, the direction of force $f$, which is assumed to be always perpendicular to the rod, will  change continuously. Decomposing the force vector along $x$ and $y$ directions \cite {fp70,n5}, motion of $C$ can then be obtained from 
\begin{eqnarray}
\label{eq:73a4}
m a_{\rm x}=-f \sin \phi=-f \sin (t^2 f \delta/2I),\\
\label{eq:73a5}
m a_{\rm y}=f \cos \phi=f \cos  (t^2 f \delta/2I).
\end{eqnarray}
Integrating with time we get
\begin{eqnarray}
\label{eq:73a6}
v_{\rm x}=-(f/m)\int_{0}^{t}\sin (t^2 f \delta/2I)\:{\rm d}t,\\
\label{eq:73a7}
v_{\rm y}=(f/m)\int_{0}^{t} \cos  (t^2 f \delta/2I)\: {\rm d}t,
\end{eqnarray}
where we have assumed the system to be at rest ({\bf v}=0) at $t=0$. Writing $k=\sqrt {f \delta/I\pi}$ and with a change of variable $t=u/k$, we can write
\begin{eqnarray}
\label{eq:73a6.1}
v_{\rm x}=\frac{-f}{mk}s(u),\\
\label{eq:73a6.2}
v_{\rm y}=\frac{f}{mk}c(u),
\end{eqnarray}
where s(u) and c(u) are  the famous Fresnel's integrals encountered in Fresnel diffraction in optics \cite {jw81} or elsewhere \cite {ch62}
\begin{figure*}
\begin{center}
\includegraphics[width=10.5cm]{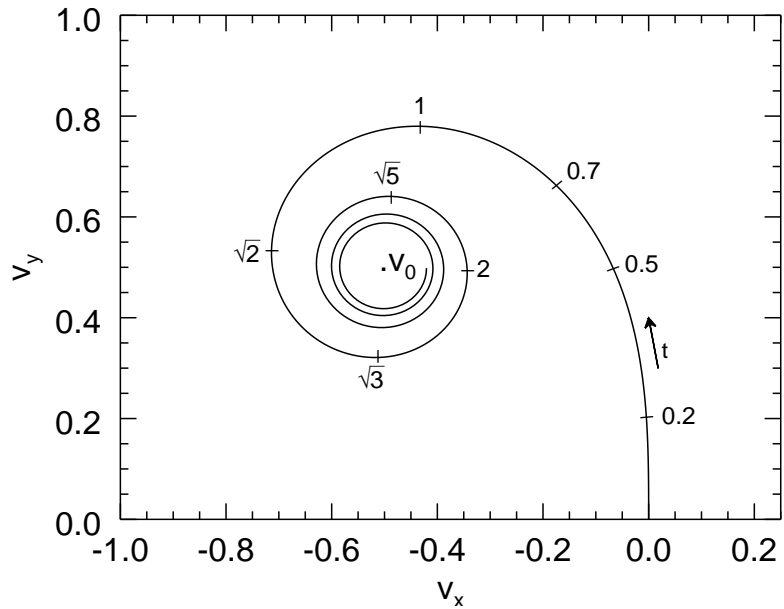}
\end{center}
\caption{Velocity vector $\bf v$ of $C$, as a function of time, describes a Cornu spiral in the $v_{\rm x}$--$v_{\rm y}$ plane, converging to ${\bf v}_0=-0.5,0.5$ (in units of $f/mk$). The  arrow indicates direction of increasing time, $t$, with tick marks on the curve showing $t$ (in units of $1/k$).}
\end{figure*}
\begin{eqnarray}
\label{eq:73a6a}
s(u)=\int_{0}^{u}\sin (\pi u^2/2 )\:{\rm d}u,\\
\label{eq:73a7a}
c(u)=\int_{0}^{u}\cos (\pi u^2/2 )\:{\rm d}u.
\end{eqnarray}
For large $t$, Eqs.~(\ref{eq:73a6.1}) and (\ref{eq:73a6.2}) yield a final constant value of $\bf v$  
\begin{eqnarray}
\label{eq:73a7.1}
v_{\rm x}=\frac{-f}{2mk}=\frac{-f}{2m}\sqrt {\frac{I \pi}{f \delta}}\:,\\
\label{eq:73a7.2}
v_{\rm y}=\frac{f}{2mk}=\frac{f}{2m}\sqrt {\frac{I \pi}{f \delta}}\:.
\end{eqnarray}

We can describe the behaviour of the velocity vector in physical terms, following \cite {n5}.
With the rotation of the rod, the direction of the force (which is applied always perpendicular to the rod) and hence that of the acceleration, will change continuously and go through cycles of $2\pi$ angle each. 
However, during each cycle, the speed of rotation will be slower at the beginning than at the end. Initially, since the force is pointing in the $y$ direction, there is a bit more velocity gained in $y$ direction. But with the rotation of the rod, as the direction of force turns towards the $-x$ direction, $C$ 
picks up velocity in that direction. These velocity gains will be substantial in the very first cycle due to the low rotation speed (Fig. 3). In later cycles, as the rod rotates faster and faster, any velocity gains during each cycle will be relatively smaller and the velocity of $C$ would soon stabilize to a constant, ${\bf v}_0$, at the centre of the Cornu spiral, as seen in Fig. 3.

As the centre of mass, $C$, of the rod will be moving with a constant final linear velocity ${\bf v_0}$, it means that the kinetic energy of translation will stabilize to a terminal value  $K_1=m v_0^2/2= \pi f l^2/48 \delta$. On the other hand 
the kinetic energy of rotation, $K_2=I \omega^2/2=6 f^2 \delta^2 t^2/m l^2$,  will be increasing indefinitely with time, with the rod spinning faster and faster about $C$. 

Assuming the centre of mass $C$ of the rod to be at rest (i.e., ${\bf v}=0$) at the origin ($x=0,y=0$) at $t=0$, the position of $C$, as a function of time, can be determined from the generic formula 
\begin{eqnarray}
\label{eq:73a8.1}
{\bf x}(t)=\int_{0}^{t}{\bf v}\:{\rm d}t={\bf v}(t)\:t-\int_{0}^{t}{\bf a}\:t\:{\rm d}t.
\end{eqnarray}
\begin{figure*}
\begin{center} 
\includegraphics[width=10.25cm]{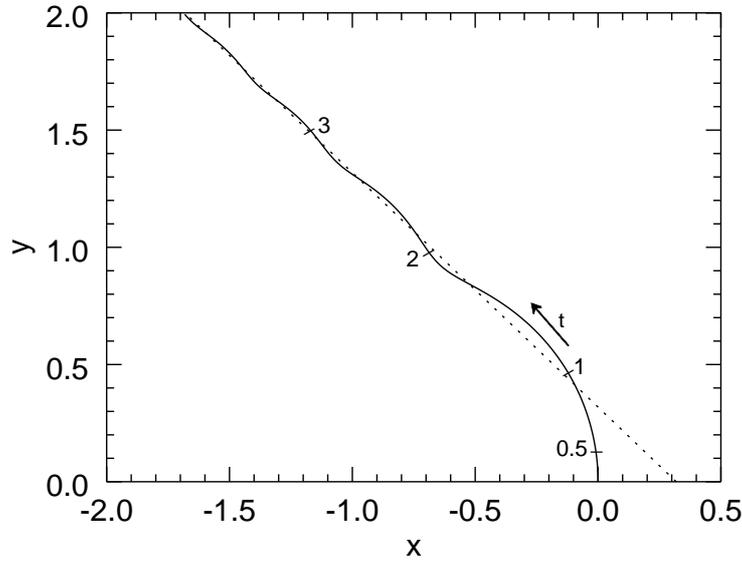}
\end{center}
\caption{Movement in $x$-$y$ plane of the centre of mass $C$, assumed initially to be at ($0,0$). It starts along a curved path which asymptotically becomes $y=-x+1/\pi$, shown as a dotted line. The distance scales are in units of $f/mk^2$. The  arrow indicates direction of increasing time, $t$, with tick marks on the curve showing $t$ (in units of $1/k$).}
\end{figure*}
Again, with a change of variable $t=u/k$ in Eq.~(\ref{eq:73a8.1}) and substituting ${\bf a}$ and ${\bf v}$ from Eqs.~(\ref{eq:73a4}), (\ref{eq:73a5}), (\ref{eq:73a6.1}) and (\ref{eq:73a6.2}), we arrive at
\begin{eqnarray}
\label{eq:73a8}
x=\frac{-f}{mk^2}\left[u\:s(u)+\frac{1}{\pi}\cos\frac{\pi u^2}{2}-\frac{1}{\pi}\right],\\
\label{eq:73a9}
y=\frac{f}{mk^2}\left[u\:c(u)-\frac{1}{\pi}\sin\frac{\pi u^2}{2}\right].
\end{eqnarray}
From Eqs.~(\ref{eq:73a8}) and (\ref{eq:73a9}), for large $t$, $C$ will follow a straight line path, with superposed initial oscillations that soon die down, as seen in Fig. 4.
\begin{eqnarray}
\nonumber
x&=&\frac{-ft}{2mk}+\frac{f}{mk^2\pi}\\
\label{eq:73a9.1}
&=&-\sqrt {\frac{\pi f l^2}{48m \delta}}\:t+\frac{l^2}{12\delta},\\
\label{eq:73a9.2}
y&=&\frac{ft}{2mk}=\sqrt {\frac{\pi f l^2}{48m \delta}}\:t,
\end{eqnarray}
and the trajectory in the $x$--$y$ plane would be $y=-x+{l^2}/{12\delta}$.

According to Ferris-Prabhu \cite{fp70}, the average motion of $C$ is along $y=-x$ (according to the conventions adopted here). It need to be pointed out that the equation (6) of Ferris-Prabhu \cite{fp70}, giving the trajectory of the centre of mass, is neither in agreement with the figure (3) given there, nor is it consistent with the initial condition that it starts from origin ($x=0,y=0$) at $t=0$. This is because the last term on the right hand side in Eq.~(\ref{eq:73a9.1}), i.e., ${l^2}/{12\delta}$, is missing from  the equation (6) of Ferris-Prabhu \cite{fp70}.
 
Actually, the velocity vector at the beginning points predominantly in the $y$ direction (Fig. 3), therefore motion of $C$ will also be initially along the $y$ direction. However, as the velocity vector reaches its final constant value $v_0$, the curved path of $C$ will asymptotically coincide with a straight line, $y=-x+{l^2}/{12\delta}$, with ${l^2}/{12\delta}$ being mostly the initial gain along the $y$ direction (Fig. 4).

If the force applied is at one end of the rod, i.e. $\delta=l/2$, then the rod will rotate about $C$ with an increasing angular speed, $\omega =6 f t/ m l$. The motion of $C$ will be still given by Eqs.~(\ref{eq:73a8}) and (\ref{eq:73a9}) but with $k=\sqrt {6f /\pi m l}$, and of course, $u=kt$.  For large $t$, the path of $C$ will follow a straight line, $y=-x+{l}/{6}$. 

However, if the force is applied at the centre of mass $C$, with $\delta=0$, then from Eqs.~(\ref{eq:73a6}), (\ref{eq:73a7}) $v_{\rm x}=0$, $v_{\rm y}=ft/m$ and from  Eqs.~(\ref{eq:73a8}) and (\ref{eq:73a9}) $x=0$, $y=ft^2/2m$. Only in such a case, the rod will then be moving linearly along the $y$ direction, albeit at an ever increasing speed, but without any accompanying rotation.
\section*{Acknowledgements} 
I acknowledge Apoorva Singal for bringing this intriguing problem to my attention, for her incessant doubts about the solutions initially suggested and for her help in the preparation of diagrams. 
{}
\end{document}